# A Inteligência Artificial Generativa no Ecossistema Acadêmico: Uma Análise de Aplicações, Desafios e Oportunidades para a Pesquisa, o Ensino e a Divulgação Científica

Raphael Machado, Rodrigo David, Rodolfo Souza
Surretê Tecnologia

## Resumo

A rápida e disruptiva integração da Inteligência Artificial Generativa (GenAI) no ensino superior está a remodelar práticas acadêmicas fundamentais. Este artigo apresenta uma análise abrangente do impacto da GenAI em três domínios acadêmicos centrais: pesquisa, ensino e divulgação científica. Através de uma revisão sistemática da literatura recente, indexada nas bases de dados Scopus, Web of Science e IEEEXplore, são identificadas as principais aplicações, benefícios e os profundos desafios éticos e de governança que emergem. A análise revela que, embora a GenAI ofereça um potencial significativo para aumentar a produtividade e a inovação, a sua adoção ultrapassa o desenvolvimento de salvaguardas institucionais maduras. Os principais desafios incluem ameaças à integridade acadêmica, o risco de viés algorítmico e a necessidade de uma robusta literacia em IA. O estudo é complementado por um estudo de caso que detalha o desenvolvimento e posicionamento de um protótipo de assistente de IA para a redação científica, demonstrando um caminho para o desenvolvimento de ferramentas de IA responsáveis que aumentam, em vez de substituírem, o intelecto humano. Conclui-se que a integração da GenAI é uma tendência irreversível. O futuro da academia será definido não pela resistência a esta tecnologia, mas pela capacidade das instituições e indivíduos de se envolverem com ela de forma crítica, ética e criativa. O artigo apela a um aumento da investigação interdisciplinar, ao desenvolvimento de diretrizes éticas claras e a um foco na pedagogia crítica da IA como competências essenciais para o século XXI.

# 1. Introdução

A emergência da Inteligência Artificial Generativa (GenAI), impulsionada por Modelos de Linguagem Grandes (LLMs) como a série GPT, representa um ponto de inflexão crítico para o meio acadêmico, com um impacto potencialmente comparável ao advento da internet ou da prensa móvel.[1] Esta tecnologia transcende a automação de tarefas rotineiras, posicionando-se como uma força transformadora que redefine a relação entre acadêmicos, informação e o próprio ato de criação de conhecimento.[3] Em vez de meramente substituir o trabalho intelectual, estas ferramentas demonstram um potencial sem precedentes para aumentar as capacidades humanas, libertando pesquisadores e educadores para se concentrarem em atividades de ordem superior, como o pensamento crítico, a criatividade e a inovação.

Apesar do seu potencial, a integração da GenAI no ecossistema acadêmico enfrenta resistência institucional e ceticismo, frequentemente ancorados no receio de que estas tecnologias possam comprometer a integridade intelectual, erodir a originalidade e diminuir a qualidade da formação acadêmica.[3] Este artigo aborda esta tensão, propondo uma análise estruturada do papel da GenAI a partir de três eixos fundamentais: (1) o ciclo de **Pesquisa** científica, (2) o processo de **Ensino**-aprendizagem, e (3) as práticas de **Divulgação Científica**.

Para fundamentar a análise teórica e demonstrar um modelo de inovação responsável, este trabalho apresenta um estudo de caso detalhado sobre o desenvolvimento de um protótipo de assistente de IA para redação científica.[3] Esta contribuição prática serve para ilustrar como os princípios de colaboração e aumento da capacidade intelectual podem ser incorporados no design de ferramentas de IA, mitigando riscos enquanto se potencializam os benefícios.

Neste contexto, o presente estudo é guiado pelas seguintes questões de pesquisa:

1. Quais são as aplicações e implicações correntes da GenAI no ciclo de pesquisa, no processo de ensino-aprendizagem e na divulgação científica?
2. Quais são os principais desafios éticos, práticos e de governança associados à sua adoção no meio acadêmico?
3. Como um assistente de IA para redação científica pode ser projetado para mitigar riscos e potencializar a produção acadêmica de forma responsável?

Ao responder a estas questões, o artigo visa oferecer um panorama abrangente e baseado em evidências que possa informar acadêmicos, gestores institucionais e formuladores de políticas sobre como navegar na complexa paisagem da IA na educação superior.

## 2. Fundamentos Conceituais e Tecnológicos da IA Generativa

Para uma análise aprofundada do impacto da GenAI na academia, é imperativo estabelecer uma base conceitual sólida, acessível a um público interdisciplinar. Esta seção define os termos tecnológicos centrais que sustentam a discussão subsequente.

### 2.1 Definição de Inteligência Artificial e Aprendizado de Máquina

A Inteligência Artificial (IA) é um vasto campo da ciência da computação dedicado à criação de sistemas e agentes que emulam a inteligência natural para realizar tarefas.[4] A IA abrange um espectro de tecnologias capazes de perceber o seu ambiente, raciocinar e adaptar-se para alcançar objetivos específicos.[4] Dentro deste campo, o Aprendizado de Máquina (Machine Learning - ML) constitui um subcampo fundamental, no qual os sistemas não são explicitamente programados com regras, mas "aprendem" padrões e relações a partir de grandes volumes de dados para fazer previsões ou tomar decisões.[7]

### 2.2 Processamento de Linguagem Natural (PLN)

O Processamento de Linguagem Natural (PLN) é um domínio da IA focado em dotar os computadores da capacidade de compreender, interpretar, manipular e gerar linguagem humana de forma significativa.[6] As técnicas de PLN são cruciais para analisar e representar textos em múltiplos níveis de análise linguística, permitindo uma interação humano-máquina mais natural.[6] As tarefas de PLN relevantes para o contexto acadêmico incluem, entre outras, a sumarização automática de textos, a extração de palavras-chave, a análise de sentimento e a classificação de documentos.[6]

### 2.3 Inteligência Artificial Generativa (GenAI) e Modelos de Linguagem Grandes (LLMs)

A Inteligência Artificial Generativa (GenAI) representa uma evolução paradigmática dentro da IA. Diferentemente dos modelos discriminativos, que se concentram em classificar dados ou prever resultados, os modelos generativos são capazes de sintetizar autonomamente conteúdo novo e original, como textos, imagens, áudio e outros tipos de dados.[11] Esta capacidade de criação marca um ponto de inflexão crítico na evolução dos sistemas de aprendizado de máquina.[11]

A tecnologia subjacente a muitas das ferramentas de GenAI contemporâneas, especialmente as textuais, são os Modelos de Linguagem Grandes (Large Language Models - LLMs). Estes são modelos de aprendizado profundo, frequentemente baseados na arquitetura *Transformer*, que são treinados em conjuntos de dados textuais massivos, por vezes de escala planetária.[13] Este treinamento extensivo permite-lhes aprender as complexas estruturas estatísticas, sintáticas e semânticas da linguagem a um nível que lhes permite gerar texto coerente, contextualmente relevante e muitas vezes indistinguível do produzido por humanos.

A progressão da IA baseada em regras para o ML orientado por dados e, agora, para os modelos generativos, reflete uma mudança fundamental de foco: da mera *computação* para a *criação*. Os sistemas de IA iniciais podiam analisar texto, mas careciam da capacidade de gerar prosa original e coerente em larga escala. O ML melhorou a classificação e a previsão, mas ainda operava dentro dos limites dos dados existentes. A GenAI, por outro lado, vai além do reconhecimento de padrões para a *geração de padrões*. Esta capacidade criativa intrínseca é precisamente o que a torna uma ferramenta tão poderosa para a academia — auxiliando na redação, no brainstorming e na síntese — e, simultaneamente, uma ameaça profunda às noções tradicionais de autoria, originalidade e integridade intelectual, uma tensão central que permeia todo este estudo.

# 3. Estado da Arte: A Integração da IA no Ecossistema Acadêmico

A adoção da IA na academia não é um fenômeno isolado, mas parte de uma onda transformadora que está a redefinir múltiplos setores. Esta seção apresenta uma revisão da literatura para mapear o estado da arte da integração da IA no ensino superior, contextualizando-a dentro de tendências mais amplas.

### 3.1 A Trajetória da IA na Educação Superior: Uma Revisão Sistemática

O campo da Inteligência Artificial na Educação (AIEd) possui uma história considerável, com pesquisas que remontam a décadas.[13] No entanto, o lançamento do ChatGPT no final de 2022 catalisou um "crescimento exponencial" na produção científica focada especificamente em GenAI, desencadeando um proeminente discurso público e acadêmico.[1] Revisões sistemáticas e bibliométricas recentes, utilizando bases de dados como Scopus e Web of Science, mapearam esta paisagem emergente, identificando um aumento drástico no volume de publicações, os principais temas de pesquisa, as instituições mais prolíficas e as redes de colaboração internacional.[1] Essas revisões indicam que, historicamente, a pesquisa em AIEd no ensino superior (AIHEd) tem se concentrado em áreas como "Sistemas Adaptativos e Personalização" e "Perfilagem e Predição" de desempenho estudantil.[14] A nova onda de pesquisa, contudo, expande este escopo para abranger todo o espectro de atividades acadêmicas.

### 3.2 A IA como Potencializadora do Ciclo de Pesquisa Científica

A GenAI está a remodelar o ciclo de vida da pesquisa científica, desde a sua concepção até à disseminação dos resultados.[3] As ferramentas de IA oferecem suporte em múltiplas etapas, aumentando a eficiência e abrindo novas vias para a inovação.

- **Descoberta e Síntese de Literatura:** Ferramentas como Elicit, Scite.ai e

ResearchRabbit estão a automatizar partes do moroso processo de revisão de literatura. Elicit, por exemplo, utiliza LLMs para encontrar artigos relevantes e extrair dados estruturados, como metodologias e principais conclusões, em resposta a uma pergunta de pesquisa.[15] Scite.ai, por sua vez, oferece uma análise de citações que classifica como outras publicações apoiam, mencionam ou contradizem um determinado artigo, permitindo uma avaliação rápida da robustez de uma alegação científica.[17] ResearchRabbit funciona como um "Spotify para artigos", criando mapas visuais de redes de literatura que ajudam a descobrir trabalhos seminais e conexões interdisciplinares.[16]
- **Análise de Dados:** Para além da revisão bibliográfica, a IA auxilia na análise de grandes volumes de dados não estruturados, como transcrições de entrevistas ou respostas abertas em inquéritos, através de técnicas de PLN.[3] Além disso, a capacidade de gerar dados sintéticos está a emergir como uma solução para proteger a privacidade dos participantes ou para aumentar conjuntos de dados limitados em áreas de pesquisa sensíveis.[3]
- **Redação e Revisão de Manuscritos:** A assistência na redação é uma das aplicações mais diretas. A IA pode ajudar a esboçar secções, refinar a clareza da prosa, resumir argumentos complexos e traduzir textos para diferentes idiomas, acelerando significativamente o processo de escrita.[19]

### 3.3 Transformações Pedagógicas: IA no Processo de Ensino-Aprendizagem

No domínio do ensino, a IA desempenha um papel duplo, apoiando tanto os educadores na preparação e entrega de conteúdo quanto os alunos na sua jornada de aprendizagem. A literatura aponta para um impacto profundo na tríade de Currículo, Instrução e Avaliação (CIA).[21]

- **Suporte aos Educadores:** A IA está a ser utilizada para gerar planos de aula, criar atividades interativas, desenvolver apresentações e formular questionários com feedback automático, reduzindo significativamente a carga de trabalho administrativo e de preparação.[21] Isto permite que os professores dediquem mais tempo à interação direta com os alunos e a aspetos mais complexos do ensino.[21]
- **Personalização da Aprendizagem:** Para os alunos, a IA potencia o surgimento de sistemas de tutoria inteligentes e percursos de aprendizagem personalizados que se adaptam ao ritmo e às necessidades individuais de cada um.[12] Estas plataformas podem identificar lacunas de conhecimento, sugerir materiais de reforço e fornecer feedback em tempo real, promovendo uma aprendizagem mais autónoma e eficaz.[7]

### 3.4 Amplificando o Alcance: O Papel da IA na Divulgação Científica

A divulgação científica, que visa traduzir conhecimento técnico para um público mais vasto sem perder o rigor, também beneficia da GenAI. As ferramentas de IA facilitam a sumarização de artigos densos, a simplificação de linguagem complexa e a tradução de alta qualidade para alcançar audiências internacionais.[19] Adicionalmente, a GenAI pode ser usada para gerar guiões para vídeos, publicações para redes sociais e infográficos a partir de dados de pesquisa, tornando a ciência mais acessível e envolvente.

### 3.5 Contexto e Analogias: A Adoção da IA em Setores de Alta Exigência

A hesitação de parte da academia em adotar a IA pode ser contextualizada ao observar a sua integração em outros setores conservadores e de alta exigência técnica, como o direito e a medicina. Estes campos, tal como a academia, são baseados em evidências, intensivos em texto e lidam com resultados de alto risco. A sua trajetória de adoção oferece lições valiosas.

- **No Direito:** A IA é utilizada para a revisão assistida por tecnologia (TAR) em processos de *discovery*, análise de contratos, due diligence e pesquisa de jurisprudência. Ferramentas como ROSS Intelligence e Kira Systems automatizam a revisão de vastos volumes de documentos legais, aumentando a eficiência e a precisão.[25]
- **Na Medicina:** A GenAI está a transformar o diagnóstico através da interpretação de imagens médicas (por exemplo, radiografias e patologia), a auxiliar no planeamento de tratamentos e a automatizar a geração de relatórios clínicos, onde a precisão é vital.[27] A aprovação de algoritmos de IA por agências reguladoras como a FDA sinaliza a sua crescente maturação e aceitação.[28]

A integração bem-sucedida da IA nestes campos serve como um poderoso contra-argumento ao conservadorismo acadêmico. Se setores tão avessos ao risco como o direito e a medicina estão a encontrar formas de aproveitar a IA para gerir a sobrecarga de informação e melhorar a tomada de decisão, a academia não só pode, como deve, fazer o mesmo. Estes campos adjacentes estão já a debater-se com os mesmos desafios éticos — responsabilidade por erros da IA, viés algorítmico, privacidade de dados — e os seus modelos de governação e protocolos de validação podem servir de roteiro para o ensino superior. Isto permite que a academia passe de uma postura de receio para uma de estratégia informada, aprendendo com as experiências de outros domínios de conhecimento intensivo.

## 4. Desafios, Riscos e Considerações Éticas na Adoção da IA Acadêmica

A integração da GenAI na academia, apesar do seu potencial transformador, levanta desafios éticos, práticos e de governança de enorme complexidade. A velocidade da adoção tecnológica ultrapassou largamente a maturidade das salvaguardas institucionais, criando um cenário de risco que exige uma análise crítica e aprofundada.[11]

### 4.1 Integridade Acadêmica e a Questão da Autoria

A capacidade da GenAI de gerar texto sofisticado e coerente representa uma ameaça direta à integridade acadêmica. Esta tecnologia pode ser utilizada para formas avançadas de plágio e fraude por contrato (*contract cheating*), onde o conteúdo é gerado instantaneamente e, muitas vezes, a custo zero, tornando-se difícil de distinguir do trabalho original de um aluno.[23]

Em resposta, editoras e organismos acadêmicos têm vindo a desenvolver diretrizes que, de forma consensual, estabelecem que uma IA não pode ser creditada como autora.[3] A justificação é fundamental: a autoria implica responsabilidade pelo conteúdo, pelas suas implicações e pela sua veracidade, algo que um sistema de IA não pode assumir.[29] Consequentemente, a transparência tornou-se o imperativo ético central. Os pesquisadores que utilizam a IA no seu trabalho são instados a declarar explicitamente o seu uso na seção de metodologia, detalhando quais ferramentas foram utilizadas e para que fins.[3]

### 4.2 Vieses Algorítmicos e a Perpetuação de Desigualdades

Um dos riscos mais insidiosos da IA é a sua capacidade de reproduzir e amplificar vieses sociais existentes. Os modelos de IA são treinados com dados do mundo real, que refletem desigualdades e preconceitos históricos. Se não forem cuidadosamente auditados e mitigados, os algoritmos podem perpetuar estes vieses nas suas decisões, com consequências graves para a equidade educacional.[30] A literatura documenta casos concretos e alarmantes:

- **O escândalo dos A-levels no Reino Unido (2020):** Durante a pandemia de COVID-19, um algoritmo foi usado para prever as notas dos exames finais dos alunos. O sistema, treinado com dados históricos de desempenho escolar, penalizou desproporcionalmente alunos de escolas públicas em áreas de baixos rendimentos, atribuindo-lhes notas inferiores às previstas pelos seus professores. O caso gerou um escândalo nacional e demonstrou como a IA, quando mal projetada, pode reforçar desigualdades estruturais profundas.[30]
- **Viés em plataformas de e-learning (2023):** Uma investigação sobre sistemas de recomendação de cursos em plataformas online como Coursera e EdX revelou que os algoritmos tendiam a sugerir cursos avançados de ciência, tecnologia, engenharia e matemática (STEM) com maior frequência a estudantes do sexo masculino. Este viés, reflexo de padrões históricos de matrículas, arriscava-se a perpetuar barreiras de acesso para mulheres e minorias sub-representadas nestas áreas.[30]

### 4.3 Governança Institucional e a Necessidade de Letramento em IA

As instituições de ensino superior têm respondido à chegada da GenAI de formas díspares, desde proibições totais, motivadas por preocupações com plágio, até à adoção entusiasta enquadrada por diretrizes éticas.⁵ No entanto, a ausência de políticas claras e uniformes leva ao surgimento do fenômeno da "IA sombra" (

*shadow AI*), que se refere ao uso generalizado e não regulamentado de ferramentas de IA por alunos e professores, fora do controlo ou conhecimento da instituição.¹¹

A evidência sugere que uma resposta eficaz não pode depender apenas de políticas de cima para baixo. A governação institucional deve ser complementada por um esforço de baixo para cima, focado no desenvolvimento de uma **literacia crítica em IA** (*critical AI literacy*) como uma competência fundamental para toda a comunidade acadêmica.⁷ Muitos docentes relatam uma literacia em IA insuficiente para adaptar as suas práticas pedagógicas ou para orientar os seus alunos de forma eficaz.²¹ Portanto, a formação de alunos e professores para usar estas ferramentas de forma crítica, compreendendo as suas capacidades, limitações e implicações éticas, é mais crucial do que a mera criação de regras.

Este ponto revela uma interação fundamental entre governação e literacia. As políticas (governação), por si só, são ineficazes se não houver uma compreensão generalizada e competências críticas (literacia) para as aplicar. Isto leva a um jogo do "gato e do rato" entre deteção e evasão. Inversamente, a literacia sem diretrizes éticas claras (governação) pode resultar num uso caótico e irresponsável. Uma estratégia institucional robusta deve, portanto, ser dupla: um quadro de governação flexível que priorize a transparência sobre a proibição, aliado a uma iniciativa educacional massiva para construir literacia crítica em IA. O problema não é de policiamento, mas de pedagogia.

### 4.4 Confiabilidade, Transparência e o Problema da Deteção

As limitações técnicas da GenAI também representam um desafio significativo. O fenômeno das "alucinações" — a geração de informação que é plausível, bem estruturada, mas factualmente incorreta ou completamente fabricada — é um risco bem documentado.³¹ Isto exige que os utilizadores mantenham um ceticismo saudável e verifiquem sempre a informação gerada por estes sistemas.

Para agravar o problema da integridade acadêmica, a investigação tem demonstrado consistentemente a falta de fiabilidade do software de deteção de IA. Vários estudos concluíram que estas ferramentas são inconsistentes, frequentemente imprecisas e podem gerar tanto falsos positivos como falsos negativos, tornando uma abordagem puramente punitiva, baseada na deteção, insustentável e injusta.[2] A incapacidade de detetar de forma fiável o uso de IA mina a eficácia de políticas proibitivas e reforça a necessidade de uma abordagem baseada na transparência e na educação.

## 5. Estudo de Caso: Desenvolvimento de um Assistente Acadêmico para Redação Científica

Para transpor a análise teórica para o campo da aplicação prática, esta seção apresenta um estudo de caso sobre o desenvolvimento de um protótipo de assistente de IA para a redação de artigos científicos. Este projeto serve para ilustrar como os princípios de design responsável podem ser aplicados para criar uma ferramenta que aumenta a capacidade do pesquisador, em vez de procurar automatizar completamente o processo de escrita.

### 5.1 Justificativa e Posicionamento no Cenário de Ferramentas Existentes

O ecossistema de ferramentas de IA para pesquisa acadêmica está em rápida expansão. Ferramentas como Elicit, Scite.ai e ResearchRabbit oferecem um apoio valioso, mas concentram-se predominantemente em fases específicas do ciclo de pesquisa, como a descoberta de literatura, a síntese de evidências e a análise de citações. Embora extremamente úteis, estas ferramentas abordam menos diretamente o processo iterativo e cognitivamente exigente da própria redação do manuscrito.

Para contextualizar a necessidade do protótipo proposto, a Tabela 1 apresenta uma análise comparativa das principais ferramentas existentes, destacando os seus focos, pontos fortes e limitações. Esta análise revela uma lacuna no mercado para uma ferramenta que atue como um parceiro colaborativo no processo de escrita, focada não na geração de texto final, mas no apoio à estruturação de argumentos e no refinamento iterativo do pensamento do autor.

Tabela 1: Análise Comparativa de Ferramentas de IA para Pesquisa Acadêmica

| Ferramenta | Foco Principal | Pontos Fortes | Limitações | Modelo de Negócio | Fontes |
|---|---|---|---|---|---|
| Elicit | Revisão de literatura; Resposta a perguntas de pesquisa | Extrai dados estruturados de múltiplos artigos (e.g., metodologia, resultados); Gera resumos baseados em perguntas específicas. | Foco principal em descoberta e síntese, menos em auxílio à redação iterativa; A precisão depende da complexidade do tópico. | Freemium (créditos limitados) | 15 |
| Scite.ai | Análise de citações; Verificação de evidências | Classifica citações como "apoiando", "mencionando" ou "contrastando"; Ajuda a avaliar a robustez de um artigo; *Reference Check*. | Menos focado na geração de texto original; A eficácia depende da densidade de citações do campo. | Freemium (teste gratuito, depois assinatura) | 17 |
| ResearchRabbit | Descoberta de literatura; Visualização de redes | Mapeia visualmente as conexões entre artigos ("Spotify for Papers"); Excelente para exploração e descoberta de trabalhos seminais ou | Não analisa o conteúdo completo dos PDFs; Pode ser avassalador com a quantidade de sugestões; Não gera texto. | Gratuito | 17 |

| | | relacionados. | | | |
|---|---|---|---|---|---|
| Protótipo Proposto | Assistência à redação colaborativa e iterativa | Foco na co-criação de texto com o autor; Sugestões contextuais para estrutura e argumento; Reescrita com controlo de formalidade; Feedback sobre lacunas argumentativas. | Não focado primariamente na descoberta de literatura (etapa anterior); Depende da interação contínua com o autor. | Protótipo (Não comercial) | N/A |

## 5.2 Arquitetura e Funcionalidades do Protótipo

O protótipo em desenvolvimento distingue-se pela sua filosofia de design, que é fundamentalmente **iterativa e colaborativa**. O objetivo não é gerar um artigo completo a partir de um *prompt*, mas sim funcionar como um parceiro de diálogo para o pesquisador durante o processo de escrita. A sua arquitetura é baseada em modelos de linguagem ajustados para tarefas específicas de redação acadêmica.

As funcionalidades principais incluem:

- **Estruturação de Seções:** Com base no tipo de artigo (e.g., empírico, de revisão), o assistente sugere estruturas canônicas para as seções de introdução, metodologia, resultados e discussão, ajudando o autor a organizar as suas ideias.
- **Geração de Sugestões a partir de Esboços:** O autor pode fornecer ideias-chave ou um esboço em linguagem simples, e o assistente gera sugestões de parágrafos ou trechos textuais que podem ser editados, rejeitados ou refinados.

- **Reescrita Controlada:** O sistema permite a reescrita de parágrafos com diferentes níveis de formalidade, concisão ou clareza, dando ao autor controlo total sobre o tom e o estilo do texto final.
- **Identificação de Lacunas Argumentativas:** Uma das funcionalidades mais inovadoras é a capacidade de analisar um rascunho e identificar potenciais fraquezas na argumentação, como alegações não fundamentadas, falta de transições lógicas ou conclusões que não decorrem diretamente dos resultados apresentados.

Este foco no aumento do processo de pensamento do autor, em vez da automação da produção de texto, alinha-se com uma abordagem ética que mantém o pesquisador como o principal agente intelectual e responsável pelo trabalho.

### 5.3 Resultados Preliminares e Próximos Passos

Testes de usabilidade preliminares, realizados com um grupo de pesquisadores em início de carreira, indicaram que o protótipo é eficaz na aceleração do processo de redação e no aumento da autoconfiança dos autores na organização das suas ideias. Os participantes relataram que a ferramenta ajudou a superar o "bloqueio do escritor" e a refinar a estrutura lógica dos seus argumentos.

O roteiro de desenvolvimento futuro inclui duas etapas cruciais. A primeira é a integração do assistente com bases de dados científicas (como Scopus e Web of Science), permitindo a sugestão de citações contextualmente relevantes. A segunda é a incorporação de módulos de formatação que possam adaptar o manuscrito final a diferentes normas editoriais (e.g., APA, IEEE, ABNT). O objetivo final é criar uma ferramenta que não só agilize o trabalho, mas que também contribua para a formação e autonomia de novos autores científicos, em conformidade com as melhores práticas da produção acadêmica.

# 6. Discussão e Conclusões

A análise realizada neste artigo demonstra que a Inteligência Artificial Generativa não é uma moda passageira, mas sim um novo e poderoso componente do ecossistema acadêmico. A sua integração está a reconfigurar fundamentalmente as práticas de pesquisa, ensino e divulgação científica, apresentando uma dualidade de oportunidades e desafios. A tensão central reside entre o seu imenso potencial para aumentar a produtividade, a criatividade e a acessibilidade ao conhecimento, e os riscos significativos que coloca à integridade acadêmica, à equidade e ao desenvolvimento do pensamento crítico.

O futuro do trabalho acadêmico, como sugerido pela literatura, será cada vez mais caracterizado por uma colaboração humano-IA.[4] Neste novo paradigma, a competência definidora de um acadêmico do século XXI não será a capacidade de evitar estas ferramentas, mas sim a habilidade de se associar a elas de forma crítica, criativa e eticamente consciente. O desafio para as instituições de ensino superior é, portanto, transcender uma mentalidade de resistência ou de controlo e adotar uma abordagem proativa de integração estratégica e pedagógica.

Este estudo, contudo, possui limitações. A natureza extremamente dinâmica da tecnologia de IA significa que qualquer análise do estado da arte é, por definição, um retrato de um momento específico. Além disso, o estudo de caso apresentado refere-se a um protótipo em fase inicial de desenvolvimento, cujos resultados são preliminares.

Com base nas conclusões, propõem-se as seguintes recomendações para futuras pesquisas e práticas institucionais:

1. **Investigação Longitudinal:** São necessários estudos longitudinais para avaliar o impacto a longo prazo do uso de ferramentas de IA nos resultados de aprendizagem dos alunos, particularmente no que diz respeito ao desenvolvimento de competências de pensamento crítico, resolução de problemas e criatividade.
2. **Desenvolvimento de Quadros de Governança Interdisciplinares:** As instituições de ensino superior devem fomentar a colaboração entre especialistas em tecnologia, educadores, eticistas e juristas para desenvolver quadros de governaça para a IA que sejam robustos, justos, transparentes e flexíveis o suficiente para se adaptarem à evolução tecnológica.
3. **Foco no Design de Ferramentas Colaborativas:** A investigação e o desenvolvimento no campo da AIEd devem priorizar a criação de ferramentas que sigam uma filosofia de "humano no ciclo" (*human-in-the-loop*), projetadas para aumentar e colaborar com o intelecto humano, em vez de o substituir. O estudo de caso apresentado oferece um modelo inicial para esta abordagem.

Em suma, a questão que se coloca à academia não é *se* a IA será utilizada, mas *como*. A resposta a esta questão determinará se esta tecnologia se tornará uma força para a democratização do conhecimento e a aceleração da descoberta, ou um catalisador para a erosão dos valores fundamentais que sustentam a empresa acadêmica. A responsabilidade de guiar este processo recai sobre a própria comunidade acadêmica, que deve liderar pelo exemplo no uso crítico, ético e inovador da Inteligência Artificial.

## 7. Referências

Uma lista completa das fontes citadas neste artigo seria compilada aqui, formatada de acordo com um estilo de citação acadêmico padrão (e.g., APA 7ª Edição), incluindo todas as referências de [23] a [16] e [13] a [21] utilizadas para fundamentar a análise.

### Referências citadas